# 229 nm UV LEDs using p-type silicon for increased hole injection

Dong Liu[1,†], Sang June Cho[1,†], Jeongpil Park[1†], Jung-Hun Seo[1,], Rafael Dalmau[2,],

Deyin Zhao[3], Kwangeun Kim[1], Munho Kim[1], In-Kyu Lee[1],

John D. Albrecht[4,*], Weidong Zhou[3], Baxter Moody[2] and Zhenqiang Ma[1, *]

[1]Department of Electrical and Computer Engineering, University of Wisconsin-Madison, Madison, WI 53706, United States

[2]HexaTech, Inc., 991 Aviation Parkway, Morrisville, North Carolina 27560, United States

[3] Department of Electrical Engineering, University of Texas at Arlington, 500 South Cooper Street, Arlington, Texas 76019, United States

[4]Department of Electrical and Computer Engineering, Michigan State University, 428 S. Shaw Lane, East Lansing, Michigan 48824, United States

[†]Authors contributed equally.

[*]Authors to whom correspondence should be addressed. Emails:

mazq@engr.wisc.edu  or jalbrech@egr.msu.edu

**Abstract**

**Ultraviolet (UV) light emission at 229 nm wavelength from diode structures based on AlN/Al$_{0.77}$Ga$_{0.23}$N quantum wells and using p-type Si to significantly increase hole injection was reported. Both electrical and optical characteristics were measured. Owing to the large concentration of holes from p-Si and efficient hole injection, no efficiency droop was observed up to a current density of 76 A/cm$^2$ under continuous wave operation and without external thermal management. An optical output power of 160 μW was obtained with corresponding external quantum efficiency of 0.027%. This study demonstrates that by adopting p-type Si nanomembrane contacts as hole injector, practical levels of hole injection can be realized in UV light-emitting diodes with very high Al composition AlGaN quantum wells, enabling emission wavelengths and power levels that were previously inaccessible using traditional p-i-n structures with poor hole injection efficiency.**

Demand for ultraviolet (UV) light emitting diodes (LEDs) is increasing due to broad applications in biological and chemical detections, decontamination, medical treatment, high density optical recording, and lithography[1-6]. The group III-nitride materials system is the most attractive candidate for UV LEDs spanning the UVA, UVB, and UVC[7-15] owing to its wide bandgap range (GaN: 3.3eV – AlN: ~6.2eV). However, as the wavelength gets shorter, the external quantum efficiency (EQE) becomes significantly degraded. Along with challenges in growth of high Al composition Al$_x$Ga$_{1-x}$N material with low defect densities, the doping concentration limitations and high ionization energy of acceptors for wide gap AlGaN render the p-side of the diode structure quite resistive and the resulting hole injection efficiency is poor. In addition, achieving an Ohmic metal contact to typical p-layers with low contact resistance remains a critical limitation to obtaining an electrically efficient DUV LED. The approach used in this work overcomes both limitations.

A variety of approaches have been resorted to circumventing the fundamental p-type doping challenges, such as polarization doping[1,5,16,17] and tunnel junctions[18-20]. Both methods require careful control of precursor fluxes for grading Al composition over the growth process, which complicates the epitaxy technique. We have reported a 237 nm UV LED using silicon as an efficient hole injector and postulated that shorter wavelength emission would be obtainable with this method [21]. In this paper, a 229 nm wavelength LED operating under continuous wave (CW) drive current is reported using higher Al composition for the multi-quantum wells (MQWs).

The UV LED structure in Fig. 1(a) was grown on a bulk AlN substrate by low pressure organometallic vapor phase epitaxy (LP-OMVPE) in a custom high-temperature reactor. As shown in Fig. 1(a), following an initial 400 nm AlN homoepitaxial layer on an AlN substrate, a Si doped (concentration: $5 \times 10^{18}$ cm$^{-3}$) 600 nm n-Al$_{0.7}$Ga$_{0.3}$N contact and electron injection layer was grown prior to the 3-period 3/6 nm Al$_{0.77}$Ga$_{0.23}$N/AlN MQWs active region. The epitaxial growth was terminated with a 20 nm Mg doped (~$1 \times 10^{18}$ cm$^{-3}$) p-GaN layer to prevent rapid oxidation of the AlN surface, the challenges of which are explained elsewhere[21]. Prior to transferring a 100 nm thick, heavily doped single-crystal p-type Si nanomembrane (NM) with doping concentration of $5 \times 10^{19}$ cm$^{-3}$, a 0.5 nm thick Al$_2$O$_3$ layer, which acts as a quantum tunnel barrier and a passivation layer, was deposited by five cycles of an atomic layer deposition (ALD) process using an Ultratech/Cambridge Nanotech Savannah S200 ALD system.

Based on the previous analysis of the surface band bending and interface induced valance offset shift between p-Si/Al$_2$O$_3$/p-GaN isotype heterojunction[21], the band alignment across the LED structure under forward bias is sketched in Fig. 1(b). With the forward bias applied, a large quantity of holes from the reservoir of the p-type Si are injected into the Al$_{0.77}$Ga$_{0.23}$N MQW region after tunneling through the thin oxide layer and drifting across the fully depleted p-type GaN layer[21]. From direct radiative recombination of holes and electrons within the MQWs, 229 nm photons were generated, which corresponds to the emission from the 3 nm Al$_{0.77}$Ga$_{0.23}$N QW energy states.

The device process flow is illustrated in Fig. 2(a)-(g). The epitaxial samples (Fig. 2(a)) were cleaned prior to the Al$_2$O$_3$ layer deposition (Fig. 2 (b)). Then the p-type Si NM was transferred, followed by a rapid thermal anneal (RTA) at 500°C for 5 minutes to increase the bonding strength between the p-type Si NM and Al$_2$O$_3$ (Fig. 2(c)). A mesa photoresist pattern was formed by photolithography image reversal, followed by a reactive ion etching (RIE) process to etch away Si and by removing p-GaN/AlN/Al$_{0.77}$Ga$_{0.23}$N/AlN MQW using an inductively coupled plasma (ICP) etcher (Fig. 2(d)). Afterward the n-type Al$_{0.7}$Ga$_{0.3}$N layer was exposed, a metal stack (Ti/Al/Ni/Au: 15/100/50/250 nm) was deposited using optical photolithography patterning, e-beam evaporation and liftoff processes. A thermal anneal at 950°C for 30 s for the cathode contact was then performed (Fig. 2(e)). Afterwards, the anode metal (Ti/Au: 15/10 nm) was then deposited (Fig. 2(f)) in the same fashion of forming the cathode metal. Each device was isolated by etching away the Si NM and further down to the homoepitaxial AlN layer (Fig. 2(g)). For optical characterizations, the AlN substrate of the devices was thinned down to 80-100 μm. A fabricated device image is shown in Fig. 2(h). The cathode and anode metal contacts were designed to be in inter-digital form to minimize the lateral current spreading-induced resistance in the n-type $Al_{0.7}Ga_{0.3}N$ contact layer. In contrast, the sheet resistance of the 100-nm p-type Si NM with $5\times10^{19}$ cm$^{-3}$ doping concentration is negligible. Since the current passing through the MQW was primarily in the region beneath the anode, the effective device area was calculated from the anode metal area and was estimated to be $1.31 \times 10^{-3}$ cm$^2$.

In order to realize substantial light output from LEDs, epitaxial layers with low threading dislocation densities and an atomically flat surface are highly desirable. Regarding the challenges, bulk AlN substrates were used in our work to grow the Al-face high Al composition AlGaN epitaxial heterostructures, reducing the dislocation density by several orders of magnitude compared with layers on non-native substrates, as the epitaxial device layers inherit the low dislocation density ($<10^4$ cm$^{-3}$) of the single-crystalline bulk AlN[21]. Reducing surface roughness of the epitaxial layers is especially critical in our scenario with a foreign membrane transfer incorporated during fabrication. Surface roughness above ~2 nm leads to non-uniform bonding between the Si-NM and GaN, which hinders carrier transport across the Si/GaN interface often results in leakage paths and reduced efficiency of the devices.

Given the crucial role of the surface roughness, we characterized the surface of the as-grown epitaxy samples and the epitaxy samples with $Al_2O_3$ layer coated and p-type Si NM bonded using both optical microscope and a Bruker Catalyst atomic force microscope (AFM). Figure 3(a) (i) shows a filtered optical microscopic image of the surface of the AlGaN sample. Figure 3(a) (ii) shows an AFM image of the epitaxial sample surface. The extracted AFM root-mean-square (RMS) surface roughness results marked on Fig. 3a (ii) were taken from a 2×2 μm$^2$ scan area. An averaged RMS roughness of 1.00 nm was measured and the images show a crack-free surface. The smooth surface allowed the high-yield (100%) transfer of Si NM to the epitaxy sample surface and also enabled the intimated contact between the Si NM and the top GaN layer.

Figure 3(b) (i) shows a filtered optical microscope image of a transferred p-Si NM on $Al_2O_3$ deposited epitaxial sample. No wrinkles or defects was induced during the p-Si NM transfer. An AFM image of the surface of the transferred p-Si NM is shown in Fig. 3 b (ii) with a 150×150 μm$^2$ surface scan area. The averaged 0.989 nm RMS surface roughness values marked in Fig. 3(b) (ii) were extracted from an 8x8 μm scan. For reference, the RMS value of p-Si NM (before transfer) was measured as 0.121 nm with 2×2 μm$^2$ AFM surface scan after NM was released from a silicon-on-insulator (SOI)[22]. It is found that the surface roughness of Si NM after transfer roughly follows that of the original roughness of the epitaxy sample surface. Since the $Al_2O_3$ layer is rather thin and conformal, it is the thinness and the softness of the Si NM that allowed the inheritance of the surface roughness.

To evaluate the electrical performance of the LEDs, current density as a function of applied voltage was measured. The measurement results are shown in Fig. 4 (a). It is seen from the linear scale plot that the LED has a typical rectifying characteristic and a turn-on voltage of about 7 V. Besides the effective voltage dropped across the MQW region for carrier injection, the p-Si/p-GaN heterojunction, non-ideal n-contact and lateral current spreading resistance in the n-AlGaN layer are responsible for part of the turn-on voltage as well. Additionally, as seen in the log scale plot the reverse current is orders lower than the forward current, which indicates that no significant leakage path, either through surface recombination or defects at the interface between Si and GaN, was formed. Electroluminescence (EL) spectra and optical power measurements were performed by coupling LED emission into a 6 inch diameter integrating sphere of a Gooch and Housego OL 770-LED calibrated spectroradiometer. Note that the LED chip did not have any light extraction fixture during EL measurements. Electrical power was supplied in constant current mode and temperature is not controlled, so the LEDs were allowed to self-heat. The linear scale of the measured EL spectra for current drive varying from 20 mA to 100 mA are shown in Fig. 4 (b). The 229 nm peak radiating from the MQWs is dominant, and with current ranging 20 mA through 100 mA the peak intensity increases correspondingly. In addition to the main peak emission, there are two discernible weaker features at 249 nm and 385 nm in the UV and blue range, respectively. The former one is from the n-$Al_{0.7}Ga_{0.3}N$ layer with 4.98 eV gap energy, which corresponds to the emitted photon wavelength of 249 nm. Since the band gap of the electron injection layer n-$Al_{0.7}Ga_{0.3}N$ is smaller than the photon energy generated by the MQWs consisting of $Al_{0.77}Ga_{0.23}N$/AlN, light emission at 229 nm is partially reabsorbed and leads to the secondary emission at 249 nm. The broad peak at 385 nm in visible range has multiple contributions, likely from the top p-GaN combined and with deep-levels in AlGaN, excited by 229 nm photons. An optical microscope image of a LED under current injection by probing are shown in the inset of Fig. 4(b) and the blue emission is clearly visible.

The optical output power was also measured without using any light extraction fixture and, as seen in Fig.4 (c), the output continuously increases with current density up to 76 A/cm$^2$, equivalent to 100 mA and eventually reaches an output power intensity of 160 µW at 24 V bias. A higher light output power is expected by using light extraction lenses with packaging. The corresponding external quantum efficiency (EQE) was calculated to be 0.027%, which could be further improved by simply thinning down the AlN substrate, which induces substantial absorption around 229 nm wavelength range due to point defects in the substrate.

In summary, we have demonstrated a 229 nm AlN/Al$_{0.77}$Ga$_{0.23}$N MQW LED with a p-type Si NM as both a p-contact and hole injection layer for high Al composition MQW structures. The light emission at 229 nm showed no significant efficiency droop up to 76 A/cm$^2$ in CW operation and without thermal management. This study provides evidence that DUV emission from electrically-injected diode structures enabled by p-type Si NM hole injection layers are a promising approach for the practical implementation of DUV LEDs and may provide a route to diode lasers in the future.

**Acknowledgement**

We acknowledge the support of the Defense Advanced Research Projects Agency (HR0011-15-2-0002) and Dr. Daniel Green.

**Figure captions**

**Figure 1. (a)** A schematic of the LED device structure. **(b)** Band diagram of the entire LED structure consisting of p-Si, p-GaN, i-Al$_{0.77}$Ga$_{0.23}$N/AlN MQWs, and n-Al$_{0.7}$Ga$_{0.3}$N contact layer under forward bias.

**Figure 2.** Fabrication process illustration of UV LED unit device. (a) Begin with a finished epitaxial sample. (b) Deposit 0.5 nm Al$_2$O$_3$ on the epitaxy sample. (c) Transfer and chemically bond 100 nm thick p-type Si NM. (d) Mesa etching down to n-type Al$_{0.7}$Ga$_{0.3}$N layer. (e) Form cathode metal contact and thermal anneal. (f) Form anode contact metal. (g) Device isolation by etching down to AlN nucleation layer. (h) An optical microscopic image of a finished LED.

**Figure 3.** Surface roughness characterizations. (a) LED epitaxy wafer before Si NM transfer: (i) A filtered optical microscopic image of cleaned epitaxy wafer. (ii) AFM images of a 20×20 μm$^2$ scan area and RMS surface roughness with 2×2 μm$^2$ AFM surface scan of the sample. (b) After p-type Si NM transfer on Al$_2$O$_3$ deposited epitaxy sample: (i) Filtered microscopic image of p-type Si NM transferred on Al$_2$O$_3$-coated epitaxy sample. (ii) AFM images of 150×150 μm$^2$ scan area and RMS surface roughness with 8×8 μm$^2$ AFM surface scan of the sample with 5×5 μm$^2$ AFM surface scan.

**Figure 4.** (a) Current density-voltage characteristics of a typical LED in linear and log scale. (b) EL spectra under different driving current densities with CW operation. Inset: an optical microscopic image of the LED diode with forward bias applied showing visible blue illumination. (c) Plot of measured light output power as a function of current density and that of the associated voltages as a function of the driving current density.

**References**

1. J. Simon, V. Protasenko, C. Lian, H. Xing, and D. Jena, Science **327**, 60 (2010).
2. J. Verma, P. K. Kandaswamy, V. Protasenko, A. Verma, H. Xing, and D. Jena, Appl. Phys. Lett. **102**, 041103 (2013).
3. Y. Taniyasu and M. Kasu, Appl. Phys. Lett. **99**, 251112 (2011).
4. J. Verma, S. M. Islam, V. Protasenko, P. Kumar Kandaswamy, H. Xing, and D. Jena, Appl. Phys. Lett. **104**, 021105 (2014).
5. S. M. Islam, K. Lee, J. Verma, V. Protasenko, S. Rouvimov, S. Bharadwaj, H. Xing, and D. Jena, Appl. Phys. Lett. **110**, 041108 (2017).
6. J. Carrano, DARPA (2005). Available at http://go.nature.com/JNBVGN.
7. J. P. Zhang, M. Asif Khan, W. H. Sun, H. M. Wang, C. Q. Chen, Q. Fareed, E. Kuokstis, and J. W. Yang, Appl. Phys. Lett. **81**, 4392 (2002).
8. Y. Taniyasu, M. Kasu, and T. Makimoto, Nature **441**, 325 (2006).
9. J. W. Orton and C. T. Foxon, Rep. Prog. Phys. **61**, 1 (1998).
10. H. Hirayama, N. Noguchi, T. Yatabe, and N. Kamata, Appl. Phys. Expr. **1**, 051101 (2008).
11. H. Hirayama, N. Noguchi, and N. Kamata, Appl. Phys. Expr. **3**, 032102 (2010).
12. T. Nishida, H. Saito, and N. Kobayashi, Appl. Phys. Lett. **79**, 711 (2001).
13. H. Hirayama, N. Maeda, S. Fujikawa, S. Toyoda, and N. Kamata, Jpn. J. Appl. Phys. **53**, 100209 (2014).
14. J. Han, M. H. Crawford, R. J. Shul, J. J. Figiel, M. Banas, L. Zhang, Y. K. Song, H. Zhou, and A. V. Nurmikko, Appl. Phys. Lett. **73**, 1688 (1998).
15. C. Pernot, S. Fukahori, T. Inazu, T. Fujita, M. Kim, Y. Nagasawa, A. Hirano, M. Ippommatsu, M. Iwaya, S. Kamiyama, I. Akasaki, and H. Amano, Phys. Status. Solidi. A. **208**, 1594 (2011).
16. O. Ambacher, B. Foutz, J. Smart, J. R. Shealy, N. G. Weimann, K. Chu, M. Murphy, A. J. Sierakowski, W. J. Schaff, L. F. Eastman, R. Dimitrov, A. Mitchell, and M. Stutzmann, J. Appl. Phys. **87**, 334 (2000).
17. Z. Peng, L. Shi-Bin, Y. Hong-Ping, W. Zhi-Ming, C. Zhi, and J. Ya-Dong, Chin. Phys. Lett. **31**, 118102 (2014).
18. S. M. Sadaf, S. Zhao, Y. Wu, Y. H. Ra, X. Liu, S. Vanka, and Z. Mi, Nano Lett **17**, 1212 (2017).
19. Y. Zhang, S. Krishnamoorthy, J. M. Johnson, F. Akyol, A. Allerman, M. W. Moseley, A. Armstrong, J. Hwang, and S. Rajan, Appl. Phys. Lett. **106**, 141103 (2015).
20. A. G. Sarwar, B. J. May, J. I. Deitz, T. J. Grassman, D. W. McComb, and R. C. Myers, Appl. Phys. Lett. **107**, 101103 (2015).
21. S. J. Cho, D. Liu, J.-H. Seo, R. Dalmau, K. Kim, J. Park, D. Zhao, X. Yin, Y. H. Jung, and I.-K. Lee, arXiv preprint arXiv:1707.04223 (2017).
22. K. Zhang, J.-H. Seo, W. Zhou, and Z. Ma, J. Phys. D: Appl. Phys. **45**, 143001 (2012).

Fig. 1

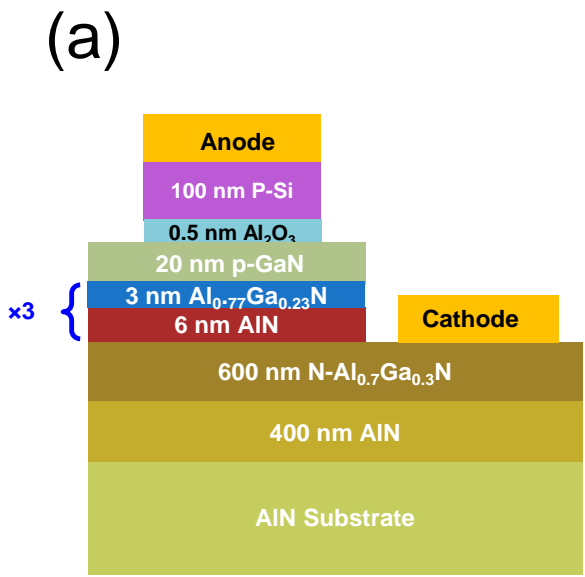 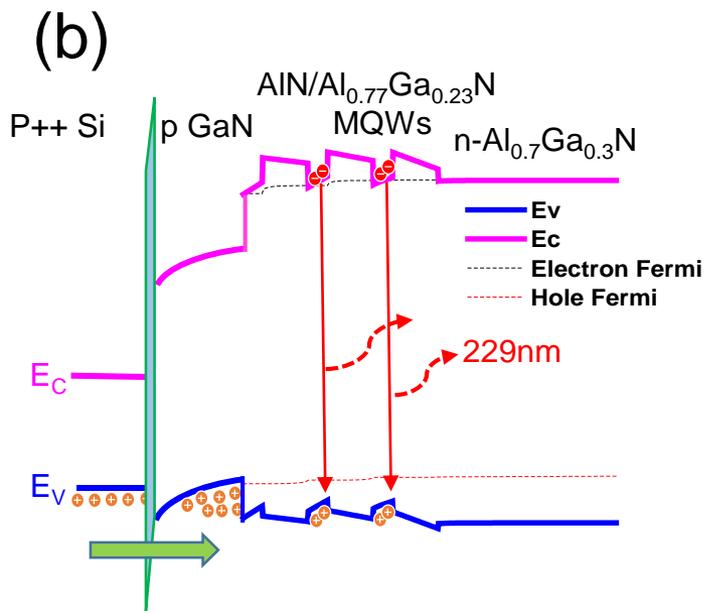

Fig. 2

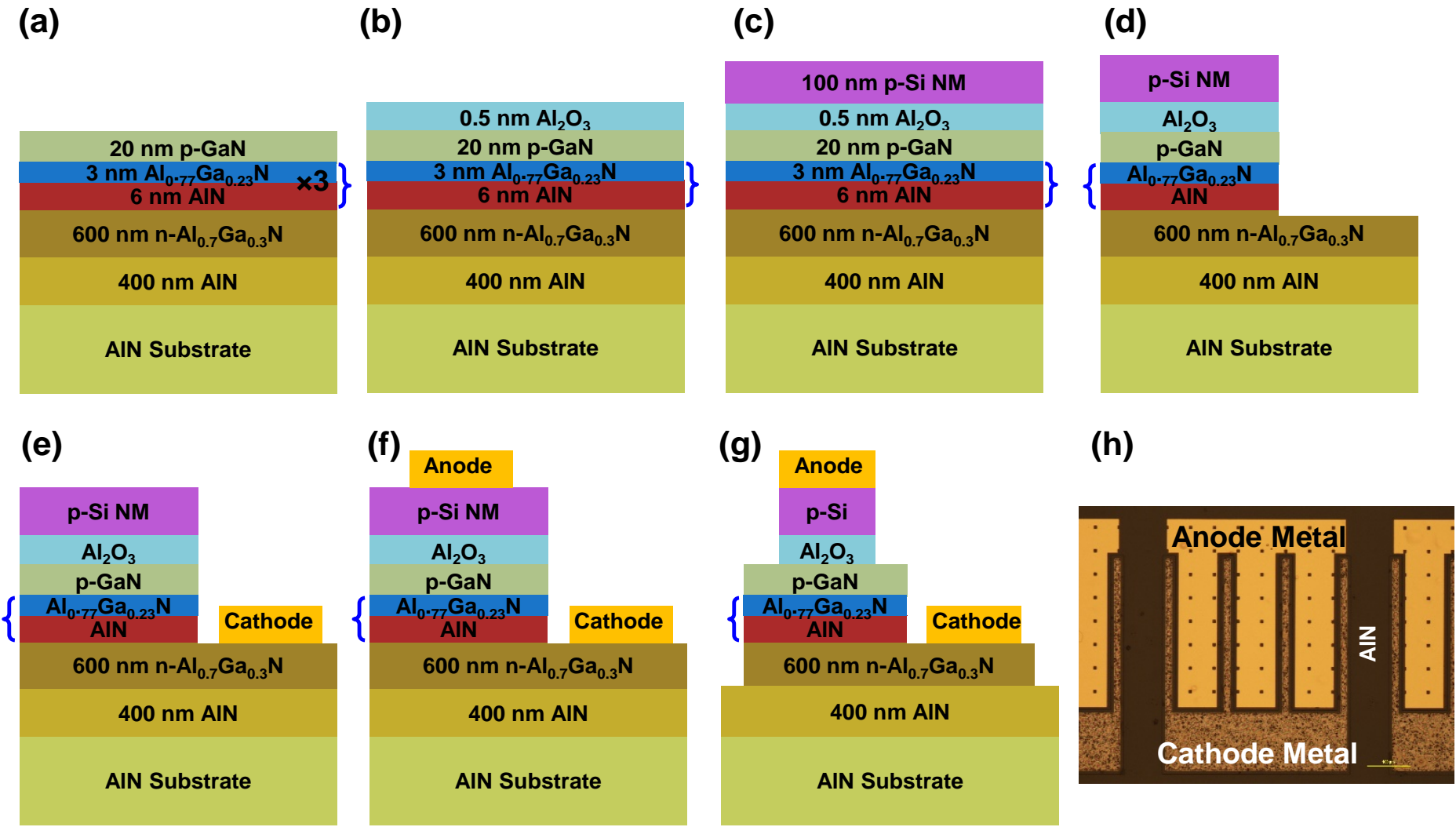

Fig. 3

(a)

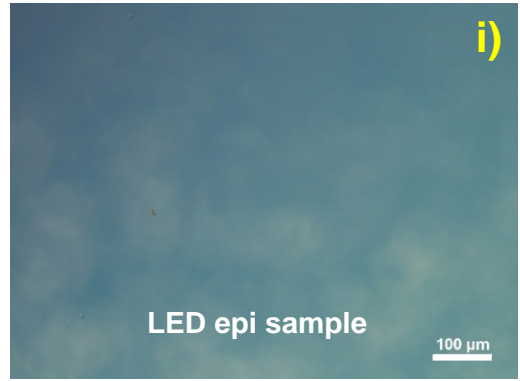 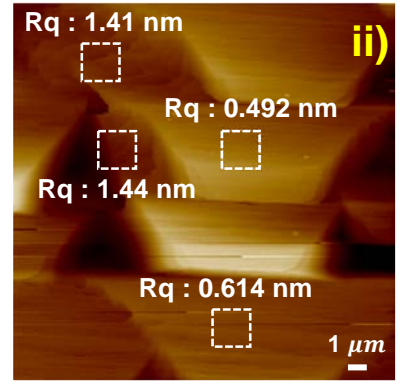

(b)

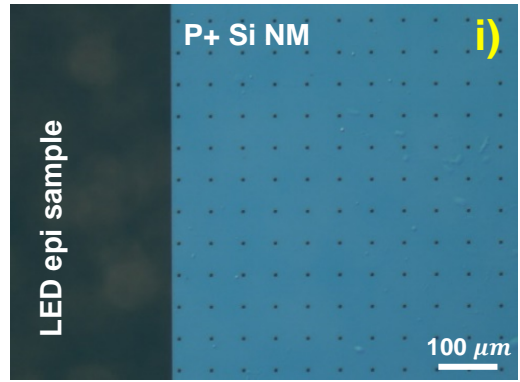 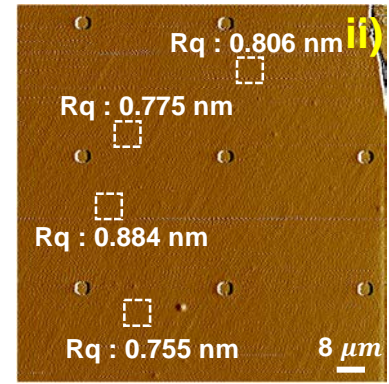

Fig. 4

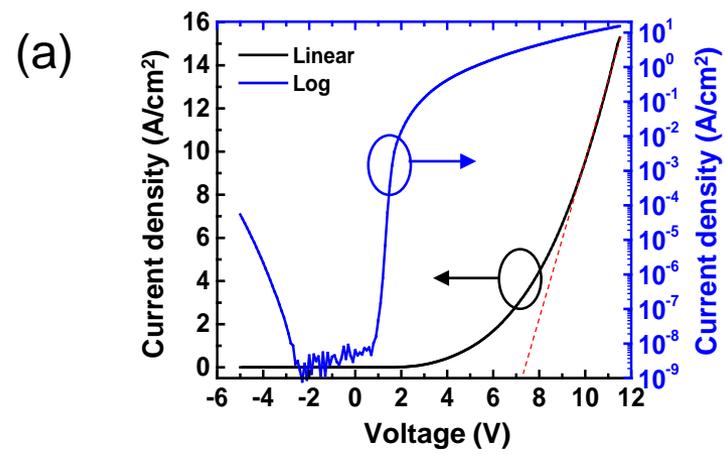

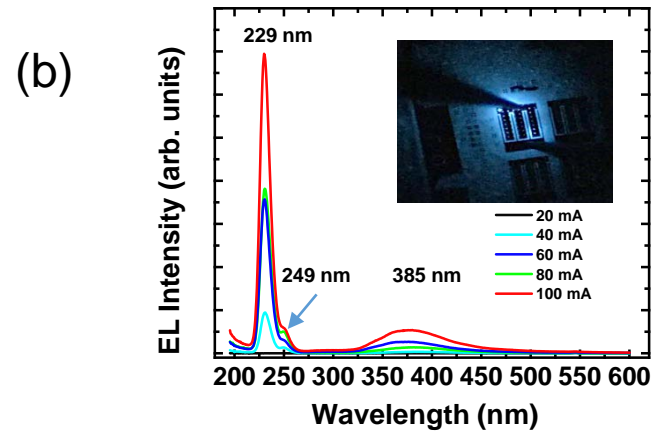

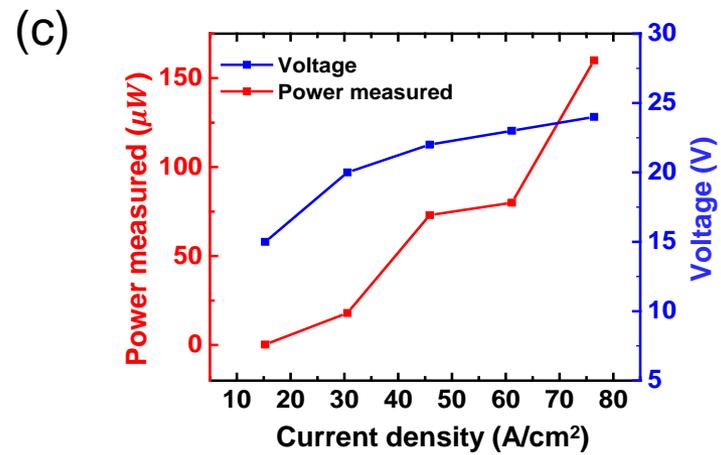